\documentclass{epl}

\title{Continuous loading of a non-dissipative atom trap}
\author{C.F. Roos, P. Cren, D. Gu\'ery-Odelin, and J. Dalibard}
\institute{
 Laboratoire Kastler Brossel\footnote{LKB is a unit\'e de recherche
 de l'Ecole Normale Sup\'erieure et de l'Universit\'e Paris 6, associ\'ee au CNRS.},
  24 rue Lhomond, F-75231 Paris Cedex 05, France }


\begin{document}

\maketitle

\begin{abstract}
We study theoretically a scheme in which particles from an
incident beam are trapped in a potential well when colliding with
particles already present in the well. The balance between the
arrival of new particles and the evaporation of particles from the
trapped cloud leads to a steady-state that we characterize in
terms of particle number and temperature. For a cigar shaped
potential, different longitudinal and transverse evaporation
thresholds can be chosen. We show that a resonance occur when the
transverse evaporation threshold coincides with the energy of the
incident particles. It leads to a dramatic increase in phase space
density with respect to the incident beam.
\end{abstract}

The trapping of atomic particles has been an invaluable tool for recent developments in atomic
physics and quantum optics. It can be performed by suddenly switching on a confining potential when
the particles are in the vicinity of its minimum. This method is used successfully to trap ions in
electromagnetic traps \cite{ions}, and neutrons \cite{neutrons} or atoms \cite{atommagnetic} in
magneto-static traps. Another way of trapping particles is to take advantage of a dissipative
mechanism, such as in a magneto-optical trap \cite{MOT}. In this way, the trap can be loaded
continuously since a friction force dissipates the excess energy of the particles and prevents them
from escaping. The loading of neutral atoms into a trap using collisions with a buffer gas also
belongs to the dissipative category \cite{buffergas}.

In this paper we investigate a different loading mechanism, in which particles are injected into a
potential well, and can be trapped by undergoing an elastic collision with one of the particles
already present in the well. After this collision, the incident particle has an energy below the
depth $U$ of the potential well and gets trapped. The excess energy is then redistributed to the
whole trapped sample via elastic collisions and it is subsequently released by the evaporation of a
trapped particle. We show that it is possible to accumulate in this way particles with an equilibrium
temperature $T\ll U/k_{\rm B}$. We also show that, for experimentally feasible conditions, a large
increase in phase space density can be achieved with respect to the one of the incident beam of
particles. Our study constitutes a realistic description of the continuous loading of a trap,
consistent with Liouville's theorem \cite{Liouville}. It has to be contrasted with the elegant, but
simplified model of \cite{Holland}, where new atoms are injected into a trap at a given energy, but
where evaporation of previously trapped atoms at the same energy is not taken into account; in that
case, the gain in phase space is only limited by ad-hoc losses, such as three-body collisions.

We develop two models in this paper. The first model deals with an
isotropic harmonic trap, assuming an evaporation criterion based
on the total energy of a particle. This allows to use simple
analytical expressions for the evaporation rate and it leads to a
proof-of-principle of the process, as long as the energy of the
injected particles is small with respect to the trap depth. The
second model is more elaborate and leads to much more spectacular
predictions for a realistic situation. It assumes an anisotropic
trap (cigar shaped potential) and it takes advantage of different
evaporation rates along the long axis of the cigar and in the
transverse directions. One can then accumulate particles in the
well even if their incident energy notably exceeds the well depth.
We conclude the paper by giving some indications on the dynamics
of the loading process.

\begin{figure}
\onefigure[scale=0.6]{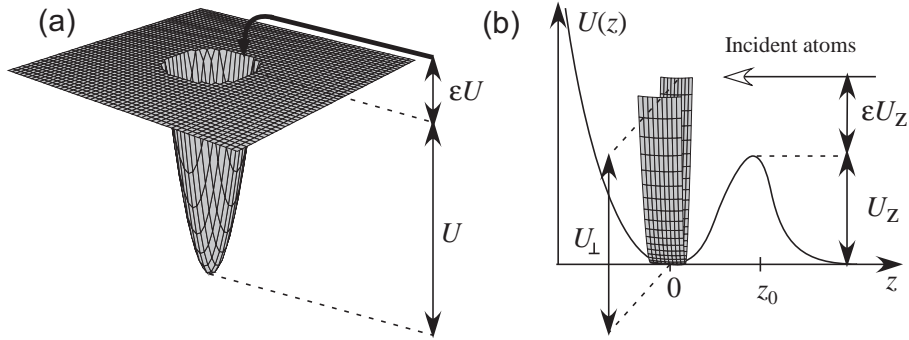} \caption{(a) First model; atoms with an energy $(1+\epsilon)U$ are
injected into a harmonic trap and can be captured by collisions with particles already present in the
trap. Trapped atoms with an energy larger than $U$ are evaporated. (b) Second model; incoming atoms
undergo collisions with trapped particles or with those reflected on the potential barrier at $z<0$.
Evaporation occurs if the longitudinal or transverse energy of a particle exceeds the evaporation
threshold $U_z$ or $U_\bot$. The phase space density of the trapped particles can be orders of
magnitude larger than the phase space density of the incoming beam.} \label{fig:scheme}
\end{figure}

For our first model, we consider an isotropic harmonic potential
with frequency $\omega$. An atomic beam with flux $\Phi$ and a
mean energy $U (1+\epsilon)$ per particle is injected into a
potential well (fig. \ref{fig:scheme}a). We assume that those
atoms undergoing collisions are trapped. Atoms are evaporated as
soon as their total (kinetic+potential) energy after a collision
exceeds the threshold $U$. The steady-state energy distribution
$P(E)$ of the trapped gas is approximated by a truncated Boltzmann
distribution with temperature $T$ \cite{Luiten}:
\[
P(E) \propto e^{-E/T}\; \theta(U-E)\ ,
\]
where $\theta$ is the Heaviside step function. For simplicity we
set the Boltzmann constant $k_{\rm B}=1$.

In order to determine $T$ and the number $N$ of trapped atoms, we
equalize the incoming and outgoing fluxes of energy and particles.
An incoming atom brings the energy $U(1+\epsilon)$. The average
energy carried away by an evaporated atom can be written $U +
\kappa\;T$, where $\kappa$ is a dimensionless coefficient which
depends on the ratio $\eta=U/T$. This coefficient can be obtained
by a simple integration over the Boltzmann equation
\cite{Luiten,note1}. From the equation
\begin{equation}
(1+\epsilon) U\;=\; U+ \kappa\; T \ \Rightarrow\
\frac{\kappa(\eta)}{\eta}=\epsilon\ , \label{energybalance}
\end{equation}
we deduce $\eta$, hence the temperature. The function
$\kappa(\eta)/\eta$ equals $1/6$ for $\eta=0$ and  decreases when
$\eta$ increases. Therefore, a steady-state exists only for values
of $\epsilon$ smaller than $1/6$. For large $\eta$, the function
$\kappa(\eta)$ tends to 1 and we obtain $\eta \simeq
\epsilon^{-1}$, i.e.
 \begin{equation}
\epsilon \ll 1 \Rightarrow T \simeq \epsilon U \ .
 \end{equation}
In other words, the equilibrium temperature is equal to the excess
energy of the incoming atoms.

We now determine the number of atoms in steady-state by comparing
the flux of atoms injected into and evaporated from the trap. We
first note that only a fraction of the incident flux $\Phi$
contributes to the feeding of the trap. Indeed, the above
description of the evaporation is only valid in the collision-less
(or Knudsen) regime, where the mean free path is much larger than
the size of the atom cloud. Therefore, most of the incoming atoms
cross the trap without any collision. To calculate the fraction
$f$ of incoming atoms which collide with a trapped atom, we assume
that all incoming trajectories pass through the center of the
trap. For the low temperatures considered here, the elastic
collisions are isotropic (s-wave regime) and they are
characterized by the total elastic cross-section $\sigma$. We find
$f=\sqrt{2\pi}\, \bar r\,n_0 \,\sigma$, where $\bar
r=\sqrt{kT/m\omega^2}$ is the size of the trapped cloud and $n_0$
the density at the center of the harmonic trap. The outgoing flux
can be calculated using classical kinetic equations. The average
collision rate is $\gamma=\sqrt{2/\pi}\,n_0\, \sigma\, \Delta v$,
where $\Delta v=\sqrt{kT/m}$. In the collision-less regime $\gamma
\ll \omega$, the probability that an atom is evaporated after a
collision is $p\simeq 2 \,\eta e^{-\eta}$ if $\eta \gg 1$
\cite{Luiten}. The balance between the incoming flux $\Phi_{\rm
in}=f\Phi$ and the outgoing flux $\Phi_{\rm out}=p\gamma N$ then
gives:
\begin{equation}
\Phi_{\rm in}=p\gamma N\  \Rightarrow \ N=\epsilon
\frac{\Phi}{\omega}\,e^{1/\epsilon}
 \ . \label{Nstat}
\end{equation}
This result is intuitive. In the regime $\epsilon^{-1}\sim \eta \gg 1$, the number of atoms in the
trap scales as $e^{1/\epsilon}\sim e^{U/T}$, which is a direct consequence of Boltzmann's law. The
number $N$ can be in principle very large. However one should keep in mind the validity criterion for
the Knudsen regime $f<1$, which imposes $e^{1/\epsilon} < U/(\Phi m \sigma \omega)$. This puts an
upper bound $N_{\rm max} \sim \epsilon U/(m\omega^2\sigma)$ on the number of trapped atoms.

Qualitatively new features arise in the second model that we now
discuss. Here, atoms are injected across the plane $z=z_0$ with a
negative velocity along the $z$ axis, and with some transverse
potential and kinetic energy (fig. \ref{fig:scheme}b). For such a
geometry, the evaporation can be either longitudinal or
transverse. The longitudinal evaporation is a direct consequence
of the loading mechanism: an atom which crosses the plane $z=z_0$
with a positive velocity is evaporated. Furthermore one can
experimentally arrange that an atom  is also evaporated when its
distance $r=(x^2+y^2)^{1/2}$ to the $z$ axis exceeds a threshold
value $r_0$. For instance, one can use a magnetic gradient to
confine the atoms transversally and use a radio-frequency wave to
flip their magnetic moment when they cross the surface of the
cylinder $r=r_0$ \cite{rfcooling}.

The possibility to control independently the transverse and
longitudinal evaporation thresholds is of particular interest if
one considers an anisotropic trap with $\omega_z \ll \omega_\bot$,
where $\omega_\bot$ and $\omega_z$ stand for the oscillation
frequencies in the $xy$ plane and along the $z$ axis,
respectively. In this case, one can reach a high-flux regime in
which the $z$ motion is in the hydrodynamic regime, i.e. $\gamma
\gg \omega_z$. All incident atoms undergo a collision with the
trapped atoms and are captured. We now study how the simple
approach presented above is modified and we derive the expected
gain in phase space density for a realistic situation.

The evaporation rate along the $z$ axis is notably reduced as
compared with the rate derived in the collision-less regime for
the same ratio $\eta_z=U_z/T$. Indeed most atoms which emerge from
a collision with an energy $E_z$ larger than the evaporation
energy $U_z=m\omega_z^2 z_0^2/2$ undergo, before reaching the
point $z_0$, another collision  which can bring their energy $E_z$
below $U_z$. Using a molecular dynamics simulation
\cite{Bird,wu,dgo1}, we have calculated (i) the probability $p_z$
that an atom reaches $z=z_0$ after a collision and is evaporated,
(ii) the average energy $U_z+\kappa_z\;T$ carried away by this
atom. For $\eta_z$ in the range from 4 to 7, we find:
\begin{equation}
\omega_z \ll \gamma\ :\qquad p_z \simeq 0.14\;e^{-\eta_z} \;
\omega_z/ \gamma \qquad \kappa_z \simeq 2.9 \ . \label{zevap}
\end{equation}

Under these operating conditions, the evaporation threshold
$U_\bot=m\omega_\bot^2 r_0^2/2$ in the $xy$ plane is a crucial
control parameter. One can set $U_\bot$ to a value larger than
$U_z$, so that the particles evaporated transversely carry an
energy $U_\bot+\kappa_\bot \,T$ notably larger than the energy
$U_z+\kappa_z \,T$ for a particle evaporated along $z$. If the
trap were operated in the collision-less regime for the $z$
motion, this would make the transverse evaporation unlikely and
inefficient. However, thanks to the reduction of $p_z$ by a factor
$\omega_z/\gamma$ with respect to the collision-less case, we can
raise $U_\bot$ to a value larger than $U_z$ and still obtain
similar probabilities $p_z$ and $p_\bot$ for longitudinal and
transverse evaporation.

The equilibrium temperature $T$ is now obtained by generalizing
(\ref{energybalance}):
 \begin{equation}
(1+\epsilon) U_z = \frac{p_z}{p_z+p_\bot} (U_z+\kappa_z T)+
\frac{p_\bot}{p_z+p_\bot} (U_\perp+\kappa_\perp T)\ .
 \label{energybalance2}
 \end{equation}
Similarly the number of atoms in steady state is obtained from the
generalization of (\ref{Nstat}):
\begin{equation}
\Phi_{\rm in}=(p_\bot+p_z)\gamma N\ . \label{Nstat2}
\end{equation}
The values of $p_z$ and $\kappa_z$ have been given in
(\ref{zevap}). We have also derived the corresponding values for
the transverse motion from a molecular dynamics simulation. For
collision dynamics ranging from the Knudsen regime ($\gamma\ll
\omega_\bot$) to the hydrodynamic regime (up to
$\gamma=5\;\omega_\bot$),  we fit our results for $\eta_\bot=8$ to
$13$ by the formula:
\begin{equation}
\quad p_\bot \simeq
2.0\;e^{-\eta_\bot}\frac{\omega_\bot}{\omega_\bot+1.4\;\gamma}\qquad
\kappa_\bot \simeq 2.0\ . \label{xyevap}
\end{equation}
The set of equations (\ref{zevap}-\ref{xyevap}) now allows us to
determine $T$ and $N$ for a given experimental situation.

As a concrete example we consider a beam of rubidium atoms
($\sigma=7.1\times10^{-16}$~m$^2$) injected in a trap with
$\omega_\bot=100\,\omega_z$ and $\omega_z/2\pi=10$~Hz. The flux is
$\Phi=10^7$ atoms/s, with an average velocity $v_i=20$~cm/s and a
velocity dispersion $\Delta v_i=4$~cm/s along each of the three
axes $x,y,z$ (temperature $T_i\sim 17\ \mu$K) \cite{cren}. This
beam is confined transversely by a harmonic potential with the
same frequency $\omega_\perp$ as in the trap. In these conditions,
the initial phase space density is ${\cal D}_i \sim 2\times
10^{-5}$. The barrier height $U_z$ in $z_0$ (see fig. 1b) is
chosen such that only atoms with an incident velocity larger than
$v_i-\Delta v_i$ are transmitted ($U_z=m(v_i-\Delta v_i)^2/2 \sim
135\ \mu$K). The flux passing in $z=z_0$ is then $\Phi_{\rm
in}=0.84\;\Phi$ and the average excess energy $\epsilon U_z$ of
the incoming atoms is $\sim 140~\mu$K (i.e. $\epsilon\sim 1$).

\begin{figure}
\onefigure[scale=0.7]{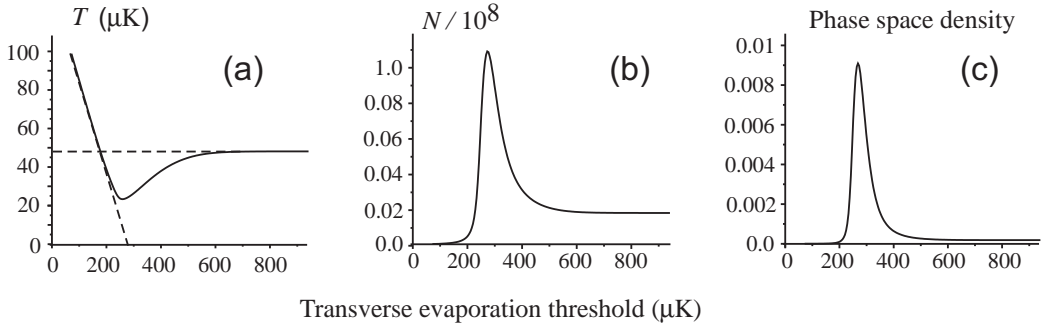} \caption{Variation of the temperature (a), of the number of trapped
atoms (b), and of the phase space density (c), as a function of the transverse evaporation threshold
$U_\bot$. The characteristics of the injected atomic beam are given in the text. The dotted lines in
(a) indicate the variation of the temperature if one neglects either longitudinal evaporation (line
with negative slope) or transverse evaporation (horizontal line).} \label{fig:gain}
\end{figure}

The temperature and the number of atoms are shown in fig.~\ref{fig:gain} as a function of the
transverse evaporation threshold. We also give the phase space density ${\cal D}$ in the trap
calculated using Boltzmann statistics. It shows a sharp maximum ${\cal D}\sim 10^{-2}$ when $U_\bot$
is close to 270~$\mu$K, in which case $p_\bot /p_z\sim 2$. The gain in phase space density with
respect to the incident beam is then $\sim 500$, while it is only 10 if one restricts evaporation to
the longitudinal direction ($U_\bot =\infty$). In this optimum regime, we find in steady-state
$\gamma\sim \omega_\bot$, $N\sim 10^8$ atoms and $T=24\ \mu$K ($\eta_z\sim 5.7$, $\eta_\bot\sim
11.4$). Note that the temperature is now much lower than the excess energy $\epsilon U_z$. This has
to be contrasted with the result of the first model where one always gets $T\geq \epsilon U$.

The spectacular resonance in the steady-state phase space density
shown in fig. \ref{fig:gain} is a general feature of our second
model. It occurs when the transverse evaporation threshold
$U_\bot$ is set to a value near the total incident energy
$(1+\epsilon)U_z$. This resonance, which allows to produce quantum
degenerate trapped clouds from incident atomic beams still far
from degeneracy, constitutes the main result of this letter.

We now discuss the characteristics of the system on the two sides
of the resonance. When $U_\bot$ is lower than the optimum value,
the evaporation is essentially transverse since $p_\bot \gg p_z$.
In this case, (\ref{energybalance2}) simplifies to give
\begin{equation}
p_\bot \gg p_z\ :\qquad T=\frac{1}{\kappa_\bot}\left(
(1+\epsilon)U_z-U_\bot)\right) \ .
 \label{Tapprox}
\end{equation}
We recover in this way the linear variation of $T$ for a small $U_\bot$ (see fig. \ref{fig:gain}).
The number of atoms, given by $\Phi_{\rm in}\sim p_\bot \gamma N$, increases as $U_\bot$ increases.
The approximation (\ref{Tapprox}) is valid until $U_\bot$ approaches $(1+\epsilon)U_z$, in which case
it would lead to $T\sim 0$. At this point the longitudinal evaporation cannot be neglected anymore,
and $p_z$ and $p_\bot$ are of the same order.

When $U_\bot$ is larger than the optimal value, the transverse
evaporation becomes inefficient and one is left with a purely
longitudinal evaporation. The temperature and the number of
trapped atoms are then independent of $U_\bot$. From
(\ref{energybalance2}-\ref{Nstat2}), one gets $T=\epsilon
U_z/\kappa_z$ (horizontal dashed line in fig. \ref{fig:gain}a) and
$N=\Phi_{\rm in}\,e^{\eta_z}/(0.14\,\omega_z)$, which is very
reminiscent of the results of the first model.

The gain $G$ in phase space is an important result of our study.
It depends on the incident flux $\Phi$ for a given loading and
trapping configuration (parameters $\epsilon$ and
$\omega_\bot/\omega_z$). For relatively small incident flux, we
find that $G$ increases with increasing $\Phi$. We consider again
the above example and we suppose that we increase $\Phi$ from
$10^7$~s$^{-1}$ to $10^8$~s$^{-1}$. Due to an increase of the
evaporation efficiency, we find that the steady-state atom number
$N$ is multiplied by 15, and that the temperature decreases by
$10\%$. The phase space gain then passes from $500$ to $900$.

The phase space gain $G$ as a function of the flux $\Phi$ saturates when the transverse motion enters
the hydrodynamic regime ($\gamma \gg \omega_\bot$). In this case, the ratio $p_\bot/p_z$ obtained
from (\ref{zevap}-\ref{xyevap}) is a function of $\eta_\bot-\eta_z$ and $\epsilon$ only. The
temperature and the ratio $N/\Phi$ are independent of the flux $\Phi$, hence a gain $G$ also
independent of $\Phi$.  Assuming $\epsilon\sim 1$ as in the example above, this maximal gain is
$G\sim 10^3$ for $\omega_\bot/\omega_z=10^2$, and $G\sim 10^4$ for $\omega_\bot/\omega_z=10^3$.

We now investigate the dynamics of the system and the time necessary to reach the steady-state of the
system around the resonance. The time evolution of the atom number $N$ and of the total energy
$3\,NT$ is given by:
\begin{eqnarray}
\frac{dN}{dt} &=& \Phi_{\rm in}-(p_z+p_\bot) \gamma N \label{dNdt} \\
\frac{d(3\,NT)}{dt} &=& \Phi_{\rm in} (1+\epsilon)U_z -p_z \gamma\, N\,(U_z+\kappa_z\,T)    -p_\bot
\gamma\, N\, (U_\bot+\kappa_\bot\,T) \ . \label{dTdt}
\end{eqnarray}
The steady-state of this set of non-linear equations corresponds to the solution of
(\ref{energybalance2}-\ref{Nstat2}). Initially, the description in terms of a thermal equilibrium
fails since no atoms are present. The loading of the trap is initiated by collisions between atoms of
the incoming and the reflected outgoing beams. In the above example, the collision rate $\gamma$
between the two beams is of the order of $\omega_z$, which ensures an efficient start of the whole
process.

The detailed study of the non-linear dynamics involved in (\ref{dNdt}-\ref{dTdt}) is outside the
scope of this letter. A simple hint on the behavior of the solutions consists in linearizing these
equations around the steady-state discussed above. Suppose for simplicity that evaporation occurs
essentially in the longitudinal direction. The two time constants of the corresponding system are
(within a numerical coefficient) $\tau_1=\tau \eta_z^2$ and $\tau_2=\tau/\eta_z^2$, where
$\tau=N_{\rm s}/\Phi_{\rm in}$ is the time required to send in the trap the number of atoms
corresponding to the steady-state number $N_{\rm s}$. Since $\eta_z\gg 1$, these two time constants
are very different from each other and the time $\tau_1$ needed to reach equilibrium exceeds by a
factor $\eta_z^2$ the ``natural" time scale $\tau$.

For the same experimental conditions as in fig. \ref{fig:gain}, we find a slow time evolution of
$N(t)$, with a time constant $\tau_1 \simeq 160$~s (to be compared with $\tau=N_{\rm s}/\Phi_{\rm
in}\sim 10$~s). By contrast, the temperature $T(t)$ reaches rapidly a value close to steady-state. We
have checked that a numerical simulation based on molecular dynamics gives values for the evolution
of $N(t)$ and $T(t)$ which differ by no more than 20\% from those deduced from the set of equations
(\ref{dNdt}-\ref{dTdt}). It is worth noting that for such long time scales losses, e.g. due to 3-body
collisions, become important and will limit in practice the gain in phase space density as in
\cite{Holland}.

To summarize, we have shown that a continuous loading of a non-dissipative trap could be achieved
with the help of evaporation. This process can be remarkably efficient in the case of an anisotropic
geometry. In this case, the incident particles can have an incident energy $\epsilon U_z$ of the
order of the entrance barrier $U_z$ and still reach a temperature $T\ll U_z$. For a realistic
configuration a gain in phase space of several orders of magnitude is possible between the incident
beam and the trapped cloud. This method can therefore be used to produce in a continuous way a
Bose-Einstein condensate. It constitutes an alternative to the proposal of \cite{Mandonnet00} based
on a continuous evaporative cooling of a guided beam. It is also complementary of the scheme recently
achieved at MIT where a succession of condensates was merged in the same trap, producing thus a
steady-state BEC in a given location \cite{Chikkatur}.

\acknowledgments C. F. Roos acknowledges support from the European Union (contract
HPMFCT-2000-00478). This work is partially supported by CNRS, Coll\`{e}ge de France, R\'egion Ile de
France, DGA, DRED and EU (TMR network ERB FMRX-CT96-0002).


\begin{thebibliography}{0}
\bibitem{note1}
One gets $\kappa(\eta)=1-P(5,\eta)/(\eta P(3,\eta)-4P(4,\eta))$, where
$P(a,\eta)\equiv
\frac{1}{\Gamma(a)}\int_0^\eta t^{a-1}e^{-t}dt$ is the incomplete Gamma
function \cite{abre}.
\end{thebibliography}
\end{document}